\documentclass[aps,pra,twocolumn,groupedaddress,showpacs,preprintnumbers,amsart]{revtex4-1}

\makeatletter
\newcommand{\figcaption}[1]{\def\@captype{figure}\caption{#1}}
\newcommand{\tblcaption}[1]{\def\@captype{table}\caption{#1}}
\makeatother

\usepackage{graphicx}
\usepackage{here}
\usepackage{amsmath}
\usepackage{amssymb}
\usepackage[mathscr]{eucal}

\begin{document}

\preprint{APS/123-QED}

\title{Rabi-coupled Countersuperflow in Binary Bose-Einstein Condensates}


\author{Ayaka Usui and Hiromitsu Takeuchi}
\affiliation{Department of Physics, Osaka City University, 3-3-138 Sugimoto, Sumiyoshi-ku, Osaka 558-8585, Japan}


\date{\today}

\begin{abstract}
We show theoretically that periodic density patterns are stabilized in two counter-propagating Bose-Einstein condensates of atoms in different hyperfine states under Rabi coupling. In the presence of coupling, the relative velocity between two components is localized around density depressions in quasi-one-dimensional systems. When the relative velocity is sufficiently small, the periodic pattern reduces to a periodic array of topological solitons as kinks of relative phase. According to our variational and numerical analyses, the soliton solution is well characterized by the soliton width and density depression. We demonstrate the dependence of the depression and width on the Rabi frequency and the coupling constant of inter-component density-density interactions. The periodic pattern of the relative phase transforms continuously from a soliton array to a sinusoidal pattern as the period becomes smaller than the soliton width. These patterns become unstable when the localized relative velocity exceeds a critical value. The stability-phase diagram of this system is evaluated with a stability analysis of countersuperflow, by taking into account the finite-size-effect owing to the density depression.
\end{abstract}


\maketitle




%

\section{INTRODUCTION}
In superfluid systems, counterflow of two interpenetrating fluid components is stable in the presence of a frictionless superfluid component. Such a flow state was first realized after the discovery of He-II, which consists of normal-fluid and superfluid components in the two-fluid model \cite{Khala}. In this system, a temperature gradient causes a counterflow of two components along the gradient, termed thermal counterflow \cite{Donne}. Thermal counterflow is an important system in the field of low-temperature physics that pertain to the visualization of quantum turbulence \cite{Paoletti, Guo}. Another interesting example is the counterflow of two superfluid components, called countersuperflow. Although countersuperflow itself must be a fundamental flow state in multi-component superfluid systems, its study received less attention until a recent experimental study on the instability of countersuperflow by Hammer et al. \cite{csEngels}. 

In their experiment \cite{csEngels}, it was found that novel soliton dynamics in quasi-one-dimension arise from the countersuperflow instability (CSI) in miscible two-component BECs \cite{Law, Yukalov, csTake, Ishino}. In this experiment, relative motion was induced by utilizing the Zeeman energy shift between the two components under a magnetic-field gradient. When relative velocity exceeds a critical value, countersuperflow becomes dynamically unstable, creating soliton in quasi-one dimensional systems. Very recently, soliton dynamics in a similar system \cite{raEngels} with an internal Josephson effect acting between two components, called Rabi coupling \cite{Rabiterm, Matthews, BdGrabi, Abad}, was observed. In the experiment \cite{raEngels}, the magnetic-field gradient induced a spatial dependency of the detuning of the coupling, which played a dominant role in soliton nucleation. This system is interesting in the sense that a countersuperflow system under Rabi coupling can be realized if the relative motion between condensates is realized in a similar manner as the earlier experiment \cite{csEngels}. It is expected that different kinds of solitons and instability developments appear in systems of Rabi-coupled countersuperflow.

Motivated by these experiments, we study theoretically Rabi-coupled countersuperflow of miscible binary condensates in quasi-one-dimensional systems. We found that the soliton patterns, which are distinctly different to those observed in Ref. \cite{csEngels, raEngels}, are stabilized in this system. In the specific limits presented here, the soliton reduces to a domain wall of a relative phase, as predicted by D. T. Son et al. \cite{Son}. This kind of structure is known to be stabilized between two vortices in the vortex-molecule structure in rotating Rabi-coupled two-component BECs \cite{vor_mole, Nitta}. However, the stability of such a structure has never been explored quantitatively, even for the quasi-one dimensional system. In this work, we present spatial profiles and stability-phase diagrams of the soliton patterns by varying the Rabi frequency, the inter-component coupling constant, and the relative velocity between the two components.

This paper is organized as follows. Section \ref{sec:sectionII} is devoted to the introduction of the stability analysis of Rabi coupled BECs and countersuperflows. In Sec. \ref{sec:sectionIII}, variational and numerical analyses are performed for the single-soliton solution in the limit of small relative velocities. Then we present the stability-phase diagram of the single-soliton. In Sec. \ref{sec:sectionIV}, the problem is generalized to the case of multi-soliton solutions with larger relative velocities. Finally, in Sec. \ref{sec:sectionV}, our results are summarized and additional discussions are made.

\section{BASIC STABILITY ANALYSES} \label{sec:sectionII}
Before discussing the soliton solution, we have to introduce bulk state, which is realized in bulk far from the soliton. First, we will formulate the stability of Rabi-coupled two-component condensate BECs without a relative velocity. Then, we will present the stability of a countersuperflow without Rabi coupling.

\subsection{Stability of Rabi-coupled condensates}
Binary BECs at zero temperature are described by the condensate wave function $\psi_{j}=\sqrt{n_{j}(x,t)}e^{i\theta_{j}(x,t)}$ $(j=1,2)$ in the Gross-Pitaevskii (GP) model \cite{PSmith}. The Lagrangian of this system under Rabi-coupling is written as
\begin{equation}
L=\int dx\; i\hbar\left(\psi_{1}^*\frac{\partial\psi_{1}}{\partial t}+\psi_{2}^*\frac{\partial\psi_{2}}{\partial t}\right)-E \label{eq:L}
\end{equation}
with the energy functional
\begin{multline}
E=\int dx\; \left\{ \sum_{j=1}^2\left(\frac{\hbar ^2}{2m}\left|\frac{\partial \psi_{j}}{\partial x}\right|^2-\mu_j|\psi_j|^2 \right. \right. \\
                                      \left. \left. +\sum_{k=1}^2\frac{g_{jk}}{2}|\psi_j|^2|\psi_k|^2
                                  \right) \right. \\
                  \left. -\frac{\hbar\Omega}{2}\left(\psi_1\psi_2^*+\psi_1^*\psi_2\right) \right\}, \label{eq:ene_den}
\end{multline}
where $m_j$  is the atomic mass, and $\mu_j$  is the chemical potential of the {\it j}th component. The coefficient $g_{jk}=2\pi\hbar^2 a_{jk}/{m_{jk}}$ of the density-density interaction is represented by the effective mass $m_{jk}=(m_j^{-1}+m_k^{-1} )^{-1}$ and the {\it s}-wave scattering length $a_{jk}=a_{kj}$ between the {\it j}th and {\it k}th components. The last term on the right hand side of Eq. (\ref{eq:ene_den}) represents the Rabi coupling \cite{Rabiterm}. We may set the Rabi frequency $\Omega$ as $\Omega\ge0$ without loss of generality.

In the presence of Rabi coupling, population transfer occurs between two components, and as such we set $m_1=m_2=m$ and $\mu_1=\mu_2=\mu$. We restrict ourselves to the case of $g_{11}=g_{22}=g>0$ and $g_{12}\ge0$, which is typically satisfied in Rabi-coupled condensates \cite{csEngels, raEngels}. Without Rabi coupling, miscible states of binary BECs are unstable for $g<g_{12}$, and the condensates undergo a phase separation. Hence, we use the non-dimensional variable
\begin{equation}
\gamma\equiv\frac{g_{12}}{g}
\end{equation}
as a characteristic parameter of this system.

The ground state is obtained by neglecting the spatial dependence of the order parameters $\psi_j$. As such, Eq. (\ref{eq:ene_den}) reduces to
\begin{multline}
E=\int dx\; \Biggl\{\sum_{j=1}^2\left(-\mu n_j +\sum_{k=1}^2 \frac{1}{2}g_{jk}n_j n_k\right) \\
                 -\hbar\Omega\sqrt{n_1 n_2}\cos\theta_-\Biggr\},
\end{multline}
where the relative phase is
\begin{equation}
\theta_-\equiv\theta_1-\theta_2.
\end{equation}
The ground state is obtained by minimizing Eq. (\ref{eq:ene_den}) with respect to $\theta_-$, $n_1$, and $n_2$. By using
\begin{equation}
n_0\equiv\frac{1}{g(1+\gamma)}\left(\mu+\frac{\hbar\Omega}{2}\right) \label{eq:n0}
\end{equation}
and
\begin{equation}
\Omega_0\equiv\frac{2g n_0}{\hbar},
\end{equation}
the ground state with $\theta_-=0$ is written as
\begin{equation}
n_1=n_2=n_0 \label{eq:bulk}
\end{equation}
for $\gamma<1+\Omega/\Omega_0$ and as
\begin{equation}
n_{1,2}=n_0\left(1\pm\sqrt{\frac{\gamma-1-\Omega/\Omega_0}{\gamma-1}}\right) \label{eq:bulk2}
\end{equation}
for $\gamma>1+\Omega/\Omega_0$ \cite{Abad}. The state (\ref{eq:bulk}) is the bulk state of our soliton solution, by assuming the condition
\begin{equation}
\frac{\Omega}{\Omega_0}<\frac{\Omega^0_c}{\Omega_0}\equiv\gamma-1. \label{eq:rc}
\end{equation}
For $\Omega=0$, the criterion (\ref{eq:rc}) reduces to that of the phase separation; $\gamma=1$. This criterion represents the global energetic stability since the analysis is based on the comparison between the energies of states (\ref{eq:bulk}) and (\ref{eq:bulk2}).

\begin{figure}[h]
\begin{center}
\includegraphics[width=8cm]{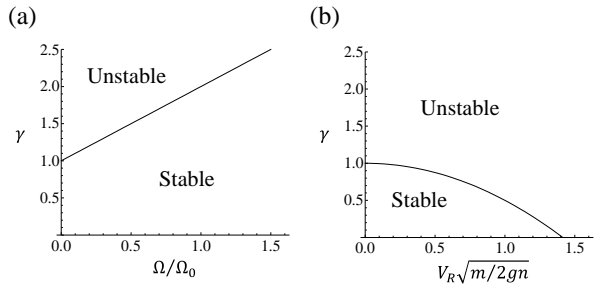}%
\caption{The stability-phase diagram of the bulk state (\ref{eq:bulk}) in Rabi coupled condensates (a) and a uniform countersuperflow (b). The phase boundaries in (a) and (b) represent Eqs. (\ref{eq:rc}) and (\ref{eq:Vc}), respectively.}\label{fig:rabi_cs}
\end{center}
\end{figure}

The local stability, the so-called linear stability, of the Rabi-coupled condensates was investigated using the Bogoliubov-de Gennes (BdG) theory \cite{Abad, BdGrabi}. Here, we investigate the linear stability around the bulk state (\ref{eq:bulk}); $\psi_j=\sqrt{n_0}$. By linearizing the equation of motion obtained from Eq. (\ref{eq:L}) with respect to a collective perturbation $\delta\psi_j(x,t)=\psi_j(x,t)-\sqrt{n_0}=u_j e^{i(qx-\omega t)}-(v_j e^{i(qx-\omega t)})^*$ and diagonalizing the linearized equations, we obtain the dispersion relations
\begin{equation}
\left(\frac{\omega}{\Omega_0}\right)^2=\frac{q^2\xi^2}{2}\left(\frac{q^2\xi^2}{2}+1+\gamma\right)
\end{equation}
and
\begin{equation}
\left(\frac{\omega}{\Omega_0}\right)^2=\left(\frac{q^2\xi^2}{2}+\frac{\Omega}{\Omega_0}\right)\left(\frac{q^2\xi^2}{2}+1-\gamma+\frac{\Omega}{\Omega_0}\right), \label{eq:BdG}
\end{equation}
where we used
\begin{equation}
\xi\equiv\frac{\hbar}{\sqrt{2mg n_0}}.
\end{equation}
Since the system is unstable for $\omega^2<0$, we find from Eq. (\ref{eq:BdG}) that the stability condition of state (\ref{eq:bulk}) is represented again by Eq. (\ref{eq:rc}). In this work, we consider the parameter region that satisfies condition (\ref{eq:rc}), as shown in Fig. \ref{fig:rabi_cs} (a).

\subsection{Stability of the countersuperflow} \label{sec:sectionIIB}
Before we discuss the stability of the Rabi-coupled countersuperflow, it is useful to demonstrate the linear stability of the countersuperflow. A stationary solution of our system is described by the time-independent GP equation
\begin{equation}
\mu_j\psi_j=\left(\frac{-\hbar^2}{2m}\frac{\partial^2}{\partial x ^2}+g|\psi_j|^2+g_{jk}|\psi_k|^2\right) \psi_j -\frac{\hbar\Omega}{2}\psi_k \label{eq:ti-GP}
\end{equation}
for $k\ne j$. For $\Omega=0$, we have the uniform solution
\begin{equation}
\psi_j=\sqrt{n_j}e^{im V_j x/\hbar}
\end{equation}
with the density
\begin{equation}
n_j=\frac{1}{g(1-\gamma^2)}\left\{\left(\mu_j-\frac{m V_j^2}{2}\right)-\gamma\left(\mu_k-\frac{m V_k^2}{2}\right)\right\} \label{eq:bulkcs}
\end{equation}
and the superfluid velocity $V_j\equiv\hbar{\partial}_x \theta_j/m={\rm const}$. We consider a countersuperflow state with a non-zero relative velocity
\begin{equation}
V_R\equiv|V_1-V_2|.
\end{equation}

Here, we will discuss the linear stability of the uniform countersuperflow with $n_1=n_2=n$, as related to the bulk state (\ref{eq:bulk}). The BdG analysis gives the dispersion relation \cite{Law, Ishino}
\begin{multline}
(\hbar \omega-V_G\hbar q)^2 =\epsilon_q^2+\epsilon_q\left(2gn+\frac{mV_R^2}{2}\right) \\
                                           \pm\sqrt{2mV_R^2(\epsilon_q+2gn)+(2g\gamma n)^2}, \label{eq:BdGcs}
\end{multline}
where $V_G\equiv(m_1n_1V_1+m_2n_2V_2)/(m_1n_1+m_2n_2)=(V_1+V_2)/2$ and $\epsilon_q\equiv\hbar^2q^2/2m$. The countersuperflow is dynamically stable when
\begin{equation}
V_R<V_c=2\sqrt{\frac{gn(1-\gamma)}{m}}. \label{eq:Vc}
\end{equation}
If $V_R>V_c$, the system becomes dynamically unstable with the non-zero imaginary part, ${\rm Im} \; \omega\ne 0$. The parameter region of condition (\ref{eq:Vc}) is represented in Fig. \ref{fig:rabi_cs} (b).

The countersuperflow can be unstable even when $V_R<V_c$ if a collective mode causes a negative-energy fluctuation $\delta \mathscr{E}=\hbar\omega\sum\nolimits_j \left(|u_j|^2-|v_j|^2\right)<0$. Here, we consider the positive norm $\sum\nolimits_j \left(|u_j|^2-|v_j|^2\right)>0$ without loss of generality. Note that dispersion (\ref{eq:BdGcs}) depends on $V_G$  as $\omega=\omega(V_G=0)+q V_G$. We have $\omega<0$ with $|V_G|>V_L={\rm min}_q[\omega(V_G=0)/q]$, and so the system is energetically unstable. This is the so-called Landau instability \cite{Khala}. This instability is physically related to the motion of the center of mass of binary condensates relative to the environment, such as an external potential or thermal excitations. The negative energy mode with $\omega<0$ is spontaneously excited and amplified due to energy dissipation. In this work, we do not consider the case where the motion of the center of mass is finite, since we are interested in the maximum stability of the Rabi-coupled countersuperflow.

For the sake of the discussion in Sec. \ref{sec:sectionIIIB}, we will also present the stability of a countersuperflow in a finite-size system. What we need to show here is the maximum wave number $q_c$  of the unstable mode, which has ${\rm Im} \; \omega\ne 0$. According to Eq. (\ref{eq:BdGcs}) we obtain
\begin{equation}
q_c=\frac{m}{\hbar}\sqrt{V_R^2-V_c^2}. \label{eq:qc}
\end{equation}
The countersuperflow is stable when $q_c$ is smaller than $1/L$, where $L$ is the system size. Note that we have ${\rm Im}\; \omega= 0$ with $V_G=0$ for $|q|\ge q_c$ and ${\rm Re}\; \omega$ arise from zero at $|q|=q_c$. This means that the system is marginally stable against the Landau instability for $q_c\sim 1/L$ in the sense that the system possesses an energetic instability for any non-zero value of $V_G$; $V_L=0$. A characteristic behavior of the energetic instability of a countersuperflow is in the momentum change $\delta J_j =\hbar q \left(|u_j|^2-|v_j|^2\right)$  owing to the instability. The instability causes a relaxation of the relative motion between the two components, and the fluctuations cause a momentum change in the opposite direction, with $|q|\gtrsim q_c>0$; $\delta J_1 \delta J_2 <0$. This characteristic behavior is revisited in Sec. \ref{sec:sectionIIIB} when we discuss the instability of the soliton.

\section{single-soliton states} \label{sec:sectionIII}
In this section, we will investigate how Rabi coupling affects a countersuperflow. In the presence of Rabi coupling, a uniform countersuperflow with $\theta_-=m V_R x/\hbar$ is not a stationary solution. We consider a relative velocity $V_R=2\pi \hbar/mL$ with a system size $L$, where the relative phase winds once through the system. In this case, a single-soliton appears as a kink of relative phase by making a density depression. Then, we discuss the stability of the single-soliton solution. The multi-soliton solution is discussed in the next section.

\subsection{The single-soliton solutions} \label{sec:sectionIIIA}
First, we will present our numerical solutions of the single-soliton before performing a detailed theoretical analysis. The solution is obtained by solving Eq. (\ref{eq:ti-GP}) numerically using imaginary time propagation (or the steepest descent method). The numerical computations were performed from an initial state $\psi_j=\sqrt{n_0}e^{i(-1)^{j+1}\pi(x/L+1/2)}$ under the Neumann boundary condition. The system size $L$ is taken to be large enough so that the system size does not affect the solution.

\begin{figure}[h]
\begin{center}
\includegraphics[width=8cm]{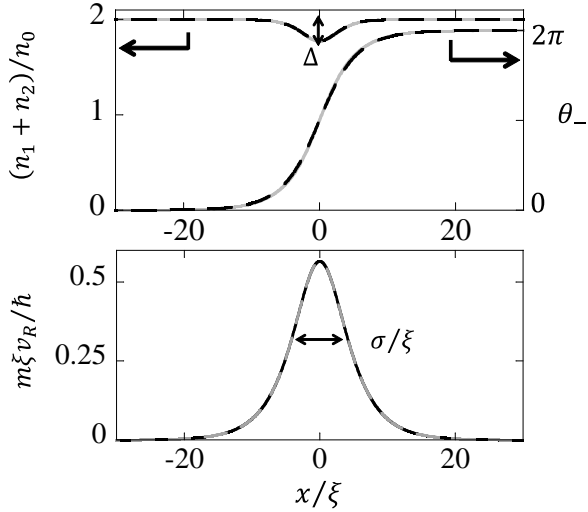}%
\caption{A typical numerical solution (black dashed curve) of the single-soliton with $\gamma=0.4$ and $\Omega/\Omega_0=0.04$. The gray dashed curves show the analytical result obtained from the variational analysis.}\label{fig:ss_profiles}
\end{center}
\end{figure}

Figure \ref{fig:ss_profiles} shows a typical profile of the solutions. The relative velocity $v_R (x)\equiv\hbar\partial_x \theta_-/m$ is localized around $x=0$ by forming a kink of relative phase $\theta_-$ and a depression in the total density $n_1+n_2$. There is no difference between the density profiles with $n_1=n_2$, and the total phase $\theta_+\equiv\theta_1 +\theta_2$  is spatially constant.

To represent the spatial profiles of the numerical results analytically, we performed a variational analysis, which is useful for simply characterizing the single-soliton solutions using parameters $\Omega$ and $\gamma$. We first determined the asymptotic behavior of $n_j (x)$ and $\theta_j (x)$ for $x \rightarrow\infty $ for constructing a variational ansatz. Equation (\ref{eq:ti-GP}) is then reduced to
\begin{equation}
0=\xi^2\frac{\partial^2 \theta_j}{\partial x ^2}+\frac{\xi^2}{n_j}\frac{\partial \theta_j}{\partial x}\frac{\partial n_j}{\partial x}-\frac{\Omega}{\Omega_0}\sqrt{\frac{n_k}{n_j}}\sin{(\theta_j-\theta_k)}, \label{eq:theta_asy}
\end{equation}
\begin{equation}
\begin{split}
\frac{2\mu}{\hbar\Omega_0}=&-\frac{\xi^2}{\sqrt{n_j}}\frac{\partial^2 \sqrt{n_j}}{\partial x ^2}+\xi^2\left(\frac{\partial \theta_j}{\partial x}\right)^2+\frac{n_j}{n_0}+\gamma\frac{n_k}{n_0} \\
&-\frac{\Omega}{\Omega_0}\sqrt{\frac{n_k}{n_j}}\cos{(\theta_j-\theta_k)}, \label{eq:n_asy}
\end{split}
\end{equation}
with $k\ne j$. From Eq. (\ref{eq:theta_asy}), we can write the asymptotic behavior of $\theta_j (x)$ for $x\to\pm\infty$ as
\begin{equation}
\theta_j (x)\sim(-1)^{j+1}\left(\frac{(1\pm1)\pi}{2}\mp e^{-|x|/\sigma_a}\right), \label{eq:theta_asyfunc}
\end{equation}
where
\begin{equation}
\sigma_a\equiv\sqrt{\frac{\hbar}{2m\Omega}}.
\end{equation}

The asymptotic form of $n_j$  is derived by inserting Eq. (\ref{eq:theta_asyfunc}) into Eq. (\ref{eq:n_asy}). With regards to the asymptotic form of the densities, there are three length scales: $\sigma_a/2$, $\xi_+\equiv\xi/\sqrt{2(1+\gamma)}$ and $\xi_-\equiv\xi/\sqrt{2(1-\gamma+\Omega/\Omega_0)}$. When $\sigma_a/2>\xi_\pm$, the asymptotic form is described as
\begin{equation}
n_j (x)\sim n_0(1-e^{-2|x|/\sigma_a}). \label{eq:n_asyfunc}
\end{equation}
The condition $\sigma_a/2>\xi_\pm$ reduces to 
\begin{equation}
\frac{4\Omega}{\Omega_0}-1<\gamma<1-\frac{3\Omega}{\Omega_0}, \label{eq:lengths}
\end{equation}
which is always satisfied for stable solitons, as shown in Sec. \ref{sec:sectionIIIB} (see also Fig. \ref{fig:stability_ss}).

We assume $n_1 (x)=n_2 (x)$ from our numerical results. If the spatial derivation of $n_j(x)$ is small so as to neglect the second term in the right hand side of Eq. (\ref{eq:theta_asy}), one obtains the sine-Gordon equation
\begin{equation}
\frac{\sigma_a^2}{2}\left(\frac{\partial \theta_-}{\partial x}\right)^2+\cos{\theta_-}=A, \label{eq:sG}
\end{equation}
where $A$ is an integration constant, and $\theta_+={\rm const.}$ because we have not considered a center of mass motion. The solutions of Eq. (\ref{eq:sG}) depend on the boundary conditions. Under the boundary condition $\theta_-\to\pi\pm\pi$ and $\partial_x \theta_-\to 0 $ for $x\to\pm\infty$, which is equivalent to $A=1$, a solution is
\begin{equation}
\theta_- (x)=4 \arctan e^{x/\sigma_a}, \label{eq:sGkink}
\end{equation}
which is called the sine-Gordon kink. This result has been obtained by neglecting the spatial derivation of $n_j$ in the limit $\gamma\to1$ \cite{Son}.

By considering the asymptotic behavior in Eq. (\ref{eq:n_asyfunc}), we constructed a variational ansatz for the density,
\begin{eqnarray}
\begin{split}
n_j (x)&=n(x) \\
&=n_0\left(1-\Delta _v {\rm sech^2} \frac{x}{\sigma_a} \right), \label{eq:density}
\end{split}
\end{eqnarray}
where $\Delta_v$ is the variational parameter. By inserting Eqs. (\ref{eq:sGkink}) and (\ref{eq:density}) into the energy (\ref{eq:ene_den}), and minimizing the energy with respect to $\Delta_v$, one obtains
\begin{equation}
\Delta_v=\frac{20\Omega/\Omega_0}{4\Omega/\Omega_0+5(1+\gamma)}.
\end{equation}
The variational ansatz proves to be a good fit to the numerical result in Fig. \ref{fig:ss_profiles}.

Here, we show the condition for applicability of our variational ansatz. Since the form (\ref{eq:sGkink}) is obtained by neglecting the second term in the right hand side of Eq. (\ref{eq:theta_asy}), the condition is satisfied for $\Delta_v\ll1$ or
\begin{equation}
\frac{\Omega}{\Omega_0}\ll\frac{5}{16}(1+\gamma)\sim1 \label{eq:neg_dn}
\end{equation}
under our assumption $0<\gamma<1+\Omega/\Omega_0$. Our approximation has no strong restriction for $\gamma$ although the limit $\gamma\to1$ is assumed in Ref. \cite{Son}.

\begin{figure}[h]
\begin{center}
\includegraphics[width=8cm]{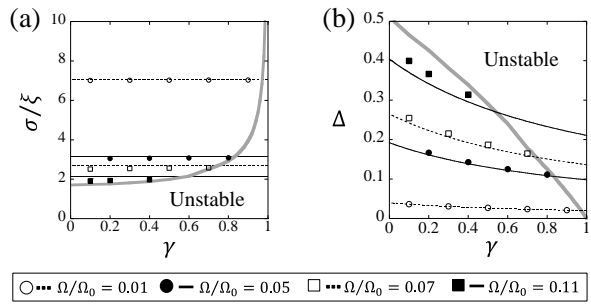}%
\caption{The soliton width $\sigma_a$ (a) and the density depression $\Delta$ (b), as functions of $\gamma$ for different $\Omega$. The numerical and analytical results are displayed with marks and lines, respectively. The gray solid lines show the stability phase boundaries of the single-soliton solution, as discussed in Sec. \ref{sec:sectionIIIB}.} \label{fig:comparison0}
\end{center}
\end{figure}

To compare the analytical results with the numerical ones in more detail, we investigated the dependences of the soliton width, $\sigma$, and density depression, $\Delta$, on $\Omega$ and $\gamma$, which are defined by
\begin{equation}
\sigma\equiv\sqrt{\frac{2}{\pi^3}\int dx\;  x^2 \left | \frac{\partial \theta_-}{\partial x} \right |}
\end{equation}
and
\begin{equation}
\Delta\equiv\frac{n_{{\rm max}}-n_{{\rm min}}}{n_{{\rm max}}}
\end{equation}
with ${\rm max}[n]\equiv n_{{\rm max}}$ and ${\rm min}[n]\equiv n_{{\rm min}}$. When we use forms (\ref{eq:sGkink}) and (\ref{eq:density}), we have $\sigma=\sigma_a$ and $\Delta=\Delta_v$. Figure \ref{fig:comparison0} shows the comparison between the analytic results $(\sigma_a, \Delta_v)$ and the numerical results $(\sigma_n, \Delta_n)$ of the variables $\sigma$ and $\Delta$. The two results coincide for small $\Omega$, consistent with condition (\ref{eq:neg_dn}), while $\Delta_v$ becomes slightly different from $\Delta_n$ for large $\Omega$. There is no numerical data in a parameter region of the plots where the single-soliton solution itself is unstable, which is revealed below.

\subsection{Stability of the single-soliton}
\label{sec:sectionIIIB}
Here we investigate the stability of the single-soliton. According to the basic stability analysis presented in Sec. \ref{sec:sectionIIB}, it is expected that the single-soliton solution is unstable if the maximum relative velocity $\sim2\pi\hbar/m\sigma_a$ at the bottom of a density depression is large enough. We show here that the stability of the single-soliton states is explained well based on the stability analysis of countersuperflow.

\begin{figure}[h]
\begin{center}
\includegraphics[width=8cm]{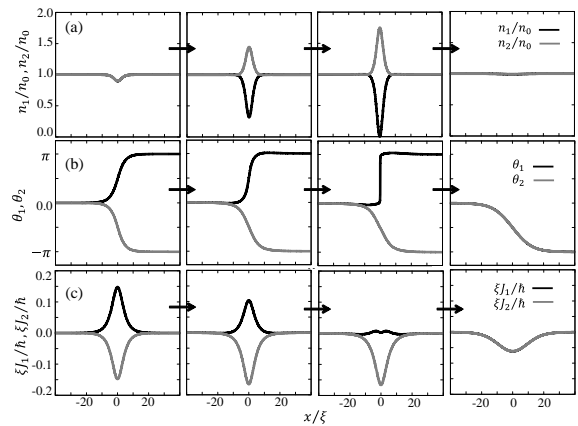}%
\caption{A typical development from left to right of densities (a), phases (b), and momentum (c)  in the numerical computation of the imaginary time propagation.}\label{fig:ima_time}
\end{center}
\end{figure}

To identify the instability, we observed the dynamics of order parameters in the imaginary time propagation of our numerical simulation. The dynamics, which effectively show a relaxation dynamic in energy-dissipative systems and does not represent an actual time development, gives us useful information, that is, what kinds of mode triggers the instability. Figure \ref{fig:ima_time} shows the instability development of the single-soliton in the imaginary-time propagation. The density difference $|n_1-n_2|$ starts to grow, and the momentum difference $|J_1-J_2|$ decreases at around $x=0$. Here, the local momentum density $J_j$ of the {\it j}th component is defined as $J_j (x)\equiv\hbar(\psi_j^* \partial_x \psi_j-\psi_j\partial_x \psi_j^*)/2i$. When the density of a component vanishes at a given point, the kink configuration in the relative phase is broken. After that, the relative velocity decays and both components flow with the same velocity.

Since the instability starts by reducing the relative momentum of the two condensates, we may expect that countersuperflow becomes unstable locally at the density depression. As a first step, we tried to apply the stability criteria for a uniform countersuperflow. However, we found that the single-soliton solution can be stable even if the maximum relative velocity $v_{{\rm max}}\equiv {\rm max}[v_R]$ at the center of the density depression exceeds the critical value $V_c=2\sqrt{g n_{{\rm min}} (1-\gamma)/m}$, which is obtained simply by using Eq. (\ref{eq:Vc}) and the minimum value $n_{{\rm min}}$, of the density. Therefore, we need to make a correction to the stability analysis in order to explain the instability of the soliton.

Since the instability occurs locally within the width $\sim\sigma_a$ in the density depression, the finite-size-effect discussed in Sec. \ref{sec:sectionIIB} should be crucial for understanding the instability criteria of the single-soliton. In this sense, the instability occurs when the soliton width $\sigma_a$ is comparable to the length $1/q_c$, where $q_c$ is the upper limit (\ref{eq:qc}). Then, we write this condition as
\begin{equation}
q_c=\frac{1}{\sigma_a}. \label{eq:condition_stability}
\end{equation}
From Eq. (\ref{eq:qc}), and using the minimum density $n_{{\rm min}}$ and the maximum velocity $v_{{\rm max}}$, the wave number $q_c$ is expressed as
\begin{equation}
q_c=\frac{m}{\hbar}\sqrt{v_{{\rm max}}^2-\frac{4g n_{{\rm min}}}{m}(1-\gamma)}.
\end{equation}
We use the variational results of the formula of $v_{{\rm max}}$ as a function of $\sigma_a$. Then, from Eq. (\ref{eq:sGkink}), we have
\begin{equation}
v_{{\rm max}}=\frac{2\hbar}{m\sigma_a}.
\end{equation}
Therefore, the criterion of instability of the single-soliton is described by
\begin{equation}
\frac{\xi}{\sigma_a}=\sqrt{\frac{2 n_{{\rm min}}}{3 n_0}(1-\gamma)}. \label{eq:condition_stability_ss}
\end{equation}
\begin{figure}[h]
\begin{center}
\includegraphics[width=8cm]{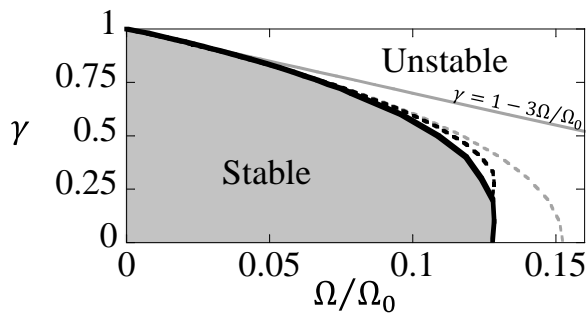}%
\caption{The stability-phase diagram of the single-soliton. The gray area surrounded by the black curve represents numerical results of the parameter region where the single-soliton is stable. The gray dashed curve shows criterion (\ref{eq:condition_stability_ss}) with $n_{{\rm min}}=n_0 (1-\Delta_v)$. The black dashed curve shows criterion (\ref{eq:condition_stability_ss}) with numerical values of $n_{{\rm min}}$. Our analysis is available below the gray line [see Eq.(\ref{eq:lengths})].} \label{fig:stability_ss}
\end{center}
\end{figure}

Figure \ref{fig:stability_ss} shows the stability-phase diagram of the single-soliton obtained from our numerical computation of the imaginary time propagation together with the plots of Eq. (\ref{eq:condition_stability_ss}). The analytical result based on the variational analysis is obtained from Eq. (\ref{eq:condition_stability_ss}) with $n_{{\rm min}}=n_0(1-\Delta_v)$. The analytical result describe well the numerical one for smaller $\Omega$, where our approximation is available. The semi-analytical results are obtained by using numerical values of $n_{{\rm min}}$ in Eq. (\ref{eq:condition_stability_ss}). The semi-analytical result coincides very well with the numerical results in Fig. \ref{fig:stability_ss}. This result shows that the stability of the single-soliton states is explained quantitatively the stability analysis of countersuperflow by taking into account the finite-size-effect.

\section{multi-soliton states} \label{sec:sectionIV}
\begin{figure}[h]
\begin{center}
\includegraphics[width=8cm]{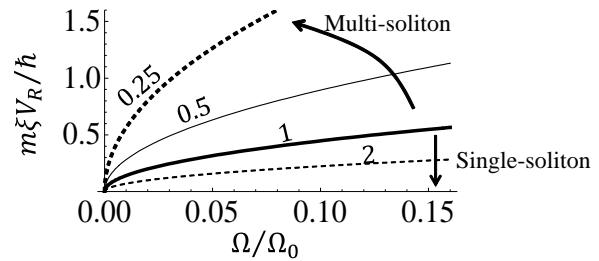}%
\caption{The $\Omega$-dependence of $V_R$ for $d/\sigma_a=0.25, 0.5, 1, 2.$} \label{fig:ds}
\end{center}
\end{figure}

In this section we discuss multi-soliton states where kinks of relative phase are so close together that stable solutions cannot be described by the single-soliton solution. The period of the soliton patterns should be $2\pi d$, with which relative phase winds
\begin{equation}
d=\frac{\hbar}{m V_R}.
\end{equation}
When $d$ is much longer than the width $\sigma_a$ of the single-solitons, the pattern is described as a periodic array of single-soliton solutions. However, this is not true when the solitons are close to each other, i.e. when $d\sim\sigma_a$. Therefore, the multi-soliton states are characterized by the ratio
\begin{equation}
\frac{d}{\sigma_a}=\sqrt{\frac{2\hbar\Omega}{m V^2_R}}. \label{eq:d_s}
\end{equation}
Under this consideration, it is straightforward to extend the analysis of the single-soliton solution in the previous section into that of the multi-soliton solution. Figure \ref{fig:ds} shows the plots of Eq. (\ref{eq:d_s}) for several values of $d/\sigma_a$. Multi-soliton states appear for smaller $\Omega$ and larger $V_R$.

\subsection{The multi-soliton solutions}
Numerical solutions of multi-solitons were obtained under the periodic boundary condition. We set the initial state for the imaginary propagation as $\psi_j=\sqrt{n_0}e^{i(-1)^{j+1}l\pi x/L}$ with integer number $l$, where the initial relative velocity was $V_R=2\pi l\hbar/mL$. 

\begin{figure}[h]
\begin{center}
\includegraphics[width=8cm]{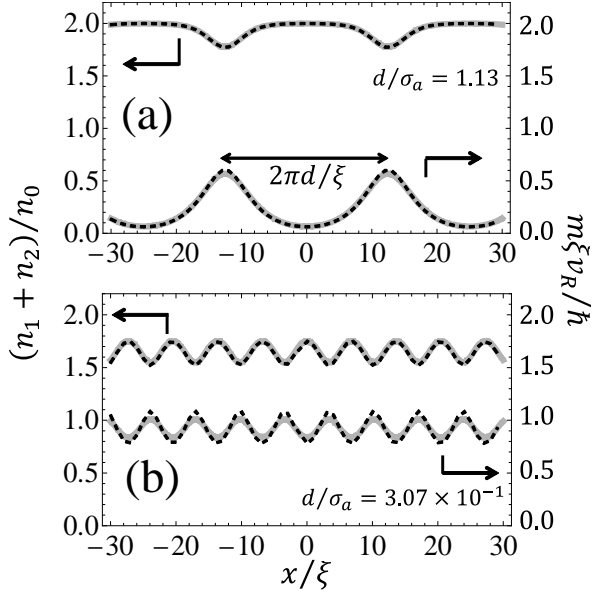}%
\caption{Numerical solutions with $\gamma=0.4$ and $\Omega/\Omega_0=0.04$. The density difference $|n_1-n_2|$ is zero, and the total phase $\theta_+$ is constant (not shown). The parameters are set as (a) $m\xi V_R/\hbar=2.51\times 10^{-1}$, where $d/\sigma_a=1.13$, and (b) $m\xi V_R/\hbar=9.22\times 10^{-1}$, where $d/\sigma_a=3.07\times 10^{-1}$.} \label{fig:ms_profiles}
\end{center}
\end{figure}

Figure \ref{fig:ms_profiles} shows typical numerical results of the multi-soliton solution. When the inter-soliton spacing becomes of order the soliton width, $d/\sigma_a\sim1$, the relative velocity is no longer localized, thus making a finite relative velocity between the density depressions in the pattern, although the density profile of each depression is still similar to that of the single-soliton solution [Fig. \ref{fig:ms_profiles}(a)]. The spatial profiles of the density and relative velocity becomes similar to a sinusoidal wave for smaller $d/\sigma_a$ [Fig. \ref{fig:ms_profiles}(b)]. Then the density becomes substantially lower than the bulk value, $n_0$.

The approximated form of the relative phase is obtained as a multi-soliton solution of the sine-Gordon equation (\ref{eq:sG}), for $A>1$. The solution is written as \cite{sG_elliptic}
\begin{equation}
\theta_-(x)=2{\rm arctan} \left[ \sqrt{1-\kappa^2} \; {\rm sc}\left[\frac{x}{\kappa\sigma_a}, \kappa \right] \right] \label{eq:sGkink_ellipitical}
\end{equation}
where
\begin{equation}
\kappa\equiv\sqrt{\frac{2}{A+1}}<1
\end{equation}
and ${\rm sc}[u,m]$ is the Jacobi elliptic function. As a natural extension of the single-soliton ansatz (\ref{eq:density}), we use the following form for the density anzats,
\begin{equation}
n(x)=n_{{\rm max}} \left(1-\Delta_v^m {\rm cn^2}\left[\frac{x}{\kappa\sigma_a}-{\rm K}[\kappa],\kappa\right]\right), \label{eq:density_ellipitical}
\end{equation}
where ${\rm K}[m]$ is the complete elliptic integral of the first kind, and $\Delta_v^m$ is a variational parameter of the multi-soliton solution. The spatial periodicity $2\pi d$ is written in terms of $\kappa$ and $\sigma_a$ as
\begin{equation}
2\pi d=2\it{K}[\kappa]\kappa \sigma_a. \label{eq:condition_d_s}
\end{equation}
Here, $n_{{\rm max}}$ represents the maximum density between the density depressions. By combining Eqs. (\ref{eq:n0}) and (\ref{eq:bulkcs}) with $V_1=-V_2=V_R/2$, $\mu_1=\mu_2=\mu$ and $v_{{\rm min}}\equiv{\rm min}[v_R]\ne 0$, the formula of $n_{{\rm max}}$ is  written approximately as
\begin{equation}
\begin{split}
n_{{\rm max}}&=\frac{1}{g(1+\gamma)}\left(\mu+\frac{\hbar\Omega}{2}-\frac{m}{8}v_{{\rm min}}^2\right) \\
&=n_0\left\{ 1-\frac{1}{4(1+\gamma)}\left(\frac{m\xi v_{{\rm min}}}{\hbar}\right)^2 \right\}. \label{eq:n_max}
\end{split}
\end{equation}
The last term in the right hand side of Eq. (\ref{eq:n_max}) shows a modification due to the non-zero relative velocity between density depressions.

The variational parameter $\Delta_v^m$ is obtained by inserting Eqs. (\ref{eq:sGkink_ellipitical}) and (\ref{eq:density_ellipitical}) into Eq. (\ref{eq:ene_den}), and then minimizing it with respect to $\Delta_v^m$. Our variational ansatz (\ref{eq:sGkink_ellipitical}) and (\ref{eq:density_ellipitical}) agree with the numerical results of the multi-soliton solutions in Fig. \ref{fig:ms_profiles}. Although our variational calculation can be inadequate for large $\Omega$, our variational calculation is available in a wide range of parameters in Fig. \ref{fig:ds}, since the multi-soliton states are stable for relatively smaller values of $\Omega$ as is disscused below.

\subsection{Stability of the multi-soliton}
Stability analysis of the single-soliton states were performed by considering the stability of countersuperflow around a density depression. For stability analysis of multi-soliton states, we have to consider an additional possibility that instability occurs in high-density regions because of nonzero relative velocity there.

To obtain a stability criterion for the former possibility, we discuss the stability of multi-soliton states in a manner similar to the stability analysis for single-soliton states. From Eq.(\ref{eq:sGkink_ellipitical}), the width of a density depression in the multi-soliton solution is represented by a length $\kappa\sigma_a$. Instability can occur when $1/\kappa\sigma_a$ equals the critical wave number (\ref{eq:qc}). The phase boundary of the instability owing to the density depressions is written in a similar form to Eq. (\ref{eq:condition_stability_ss}) as
\begin{equation}
\frac{\xi}{\kappa\sigma_a}=\sqrt{\frac{2n_{{\rm min}}}{3 n_0}(1-\gamma)}. \label{eq:gamma_c1}
\end{equation}
This criterion reduces to the stability criterion (\ref{eq:condition_stability_ss}) in the single-soliton limit $d/\sigma_a \to \infty$ with $\kappa\sigma_a\to\sigma_a$. In the opposite limit $d/\sigma_a\to 0$ with $d$ fixed, density depressions are negligibly small with $\Omega/\Omega_0 \to 0$. Then, the criterion (\ref{eq:gamma_c1}) must be reduced to the criterion of uniform countersuperflow obtained from Eq. (\ref{eq:Vc}) with $V_c=v_{\rm max}$ and $n=n_{\rm min}$. However, the criterion (\ref{eq:gamma_c1}) in the limit does not give a correct results owing to the asymptotic behavior $\kappa \sigma_a \to 2d$ for $d/\sigma_a \to 0$, with which the finite-size-effect remains even for $\Omega=0$ without density depressions.

This inconsistency is recovered by the additional possibility that instability occurs in high-density regions, where the density takes the maximum value, $n=n_{\rm max}$. Supposing that instability occurs when the relative velocity $v_{\rm min}$ in high-density regions exceeds the critical value (\ref{eq:Vc}) with $n=n_{\rm max}$, the stability criterion for small $\Omega/\Omega_0$ is written as
\begin{equation}
\frac{\xi}{\kappa\sigma_a}=\sqrt{\frac{n_{{\rm max}}}{2 n_0}(1-\gamma)}. \label{eq:gamma_c2}
\end{equation}
This criterion reduces to that of uniform countersuperflow consistently in the uniform countersuperflow limit, $d/\sigma_a\to 0$ and $\Omega/\Omega_0\to 0$.
\begin{figure}[h]
\begin{center}
\includegraphics[width=8.5cm]{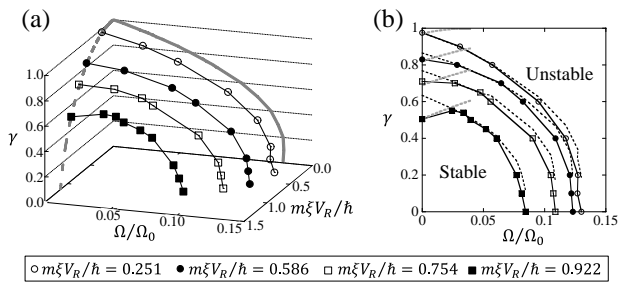}%
\caption{(a) The stability-phase diagram of the Rabi-coupled countersuperflow. The gray solid curve shows the stability-phase boundary in Fig. \ref{fig:stability_ss}. The gray dashed curve represents the stability criterion (\ref{eq:Vc}) of uniform countersuperflow. (b) The two dimensional plot of (a). The black dashed curves show the criterion (\ref{eq:gamma_c1}) with numerical data of $n_{{\rm min}}$. The gray dashed curves show the criterion (\ref{eq:gamma_c2}).} \label{fig:stability_ms}
\end{center}
\end{figure}

Stability-phase diagram of our system is summarized in Fig. \ref{fig:stability_ms}. Numerical plots are consistent with the semi-analytical estimations based on Eqs. (\ref{eq:gamma_c1}) and (\ref{eq:gamma_c2}). The analytical curve of Eq. (\ref{eq:gamma_c2}) crosses the semi-analytical curve of Eq. (\ref{eq:gamma_c1}) in the left side of the phase diagram in Fig. \ref{fig:stability_ms} (b). This structure is also consistent with the numerical plots. These results show that the stability of this system is described totally by the stability analysis of countersuperflow.

\section{SUMMARY AND DISCUSSION} \label{sec:sectionV}
We studied the Rabi-coupled countersuperflow of binary Bose-Einstein condensates in quasi-one-dimension. The variational formulas provide a good description of the stationary states of the single-soliton [Eqs. (\ref{eq:sGkink}) and (\ref{eq:density})], and the multi-soliton solutions [Eqs. (\ref{eq:sGkink_ellipitical}) and (\ref{eq:density_ellipitical})] for small Rabi frequencies. By taking into account the finite-size-effect due to density depressions in the soliton solutions, the stability analysis of countersuperflow is applicable when explaining the stability-phase diagram of the Rabi-coupled countersuperflow, Fig. \ref{fig:stability_ss} for single-soliton states, and Fig. \ref{fig:stability_ms} for multi-soliton states.

These solitons will be observed in experiments as a density contrast in atomic clouds. Therefore, the parameter dependence of the density depression in Fig. \ref{fig:ss_profiles} is an important benchmark for observing solitons experimentally. For a density contrast of higher than 5\%, $\Delta\gtrsim0.05$, $\gamma=g_{12}/g$ must be smaller than $\sim0.9$. This condition will be achieved in future experiments by utilizing Feshbach resonance \cite{Fesh}. Even for $\gamma\sim 1$, we may expect a unique behavior of this system; e.g. instability development can be distinct from those of countersuperflow instability without Rabi-coupling \cite{csEngels, csTake, Ishino} and Rabi-coupling-dominant pattern formation \cite{raEngels}. Studies of instability development provides an interesting framework for future investigations.

\begin{acknowledgments}
We are grateful to P. Engels for useful discussion. This work was supported by KAKENHI from JSPS (Grants No. 25887042 and No. 26870500). This work was also supported by the Topological Quantum Phenomena (Grants-in-Aid No. 22103003) for Scientific Research on Innovative Areas from the Ministry of Education, Culture, Sports, Science and Technology (MEXT) of Japan.
\end{acknowledgments}

\bibliography{apstemplate}

\end{document}